# Viscosity of a nucleonic fluid

Aram Z. Mekjian

Department of Physics and Astronomy, Rutgers University, Piscataway, NJ 08502*

## Abstract

The viscosity of nucleonic matter is studied both classically and in a quantum mechanical description. The collisions between particles are modeled as hard sphere scattering as a baseline for comparison and as scattering from an attractive square well potential. Properties associated with the unitary limit are developed which are shown to be approximately realized for a system of neutrons. The issue of near perfect fluid behavior of neutron matter is remarked on. Using some results from hard sphere molecular dynamics studies near perfect fluid behavior is discussed further.

## I. Introduction

Viscosity is of interest in many different areas of physics which include atomic systems, nuclear matter, neutron star physics, low energy to relativistic energy heavy ion collisions and at the extreme end string theory. Viscosity plays a role in collective flow phenomena in medium and relativistic heavy ion collisions (RHIC) where viscosity resists flow in such collisions. Ref. [1,2] are early theoretical studies of viscosity and flow in hadronic systems. Surprisingly, in relativistic heavy ion collisions, the quarks and gluons act as a strongly coupled liquid [3,4,5] with low viscosity rather than a nearly ideal gas of asymptotically free particles with high viscosity. Low viscosity fluids which interact strongly are called nearly perfect fluids. Some recent experimental results for RHIC physics can be found in Ref. [6] and some overviews are in Ref. [7,8]. String theory has put a small lower limit on the ratio of shear viscosity $\eta$ over entropy density $s$ given by $\eta/s \geq \hbar/(4\pi k_B)$ [9] with $k_B$ the Boltzmann constant. The string theory result has generated considerable interest regarding questions concerning strongly correlated systems and viscosity. The focus of the present paper is on the viscosity of a low to moderate energy systems of interacting nucleons. The nucleonic system also offers a place to study the unitary limit of thermodynamic functions and its role in viscosity.

Properties of interacting quantum degenerate Fermi gases and the unitary limit were first observed in cold atoms [10-12]. Feshbach resonances are used to study the strong coupling crossover from a Bose–Einstein condensate of bound pairs to a BCS superfluid state of Cooper pairs. A remarkable aspect of strongly interacting Fermi gases is a universal behavior which occurs when the scattering length is very large compared to the interparticle spacing. In this unitary limit, properties of a heated gas are determined by the density $\rho$ and temperature $T$, independent of the details of the two body interaction. Early theoretical discussions of degenerate Fermi systems at infinite scattering length can be found in Ref. [13,14]. At temperature $T = 0$, the Fermi energy $E$ of a strongly interacting Fermi gas differs from the Fermi energy $E_F$ of a non-interacting Fermi gas by a universal factor $\xi$ with $E = \xi E_F$. Accounting for this difference in nuclear systems is referred to as the Bertsch challenge problem [15]. Initial work on this problem was done by Barker [16] and latter Heiselberg [17]. A Monte Carlo numerical study of the unitary limit of pure neutron matter is given in Ref. [18]. Analytic studies of pure neutron systems can be found in the extensive work of Bulgac and collaborators [19-21]. In these studies the dimensionless factor $\xi \approx 0.4$. The behavior of various thermodynamic functions at finite temperature in the unitary limit and the role of Feshbach resonances in nucleonic systems were given in Ref. [22,23] extending early work given in Ref. [24,25].

Part of present paper is an extension of an earlier work [1] in which the viscosity of hadronic matter and associated flow were studied using a relaxation time approximation to the Boltzmann equation and also a Fokker-Planck description. A relaxation time approach also appears in Ref. [26,27] for the viscosity of a trapped Fermi gas in an oscillator well near the unitary limit of large scattering length. The approach taken in the present paper involves both classical and a quantum approaches to the viscosity which are discussed using a Chapman-Enskog description [28,29]. The discussion of viscosity from various potential shapes are developed which include a repulsive hard core potential and an attractive square well potential. A study of viscosity using a delta-shell potential can be found in Ref. [30,31]. The focus here will be on a pure one component system- such as in a gas of neutrons. One important feature of the interaction between nucleons is the very large scattering length $a_{sl}$ and its association with the unitary regime. The unitary limit of the viscosity is examined for this system. A study of Feshbach resonances and the second virial coefficient in atomic systems is given in ref. [32,33]. Viscosity also appear in the damping of giant resonances in nuclear physics [34] and atoms in a trap [35]. An uncertainty relation governing $\eta$ over $s$ was first realized by Danielewicz and Gyulassy [2]. An overview of viscosity and near perfect fluid behavior in several systems is given in Ref. [36] and further discussions can be found in Ref. [37-39].

## II. Classical and quantum approaches to the viscosity.
## II.A. Simplified description of viscosity

The coefficient of shear viscosity $\eta$ is defined by $F / A = \eta(du_x / dy)$ for viscous flow between two parallel plates. One plate is fixed while the other is being moved with a tangential force per unit area $F / A$ applied to it and $(du_x / dy)$ is the velocity gradient of the fluid flow between the plates. The $y$ axis is perpendicular to the two plates and $u_x$ is the fluid velocity parallel to the plates. A simplified textbook discussion of the viscosity [40] relates the viscosity to the transport of momentum across a surface and gives

$$\eta = \frac{1}{3} \rho m \hat{v} l_\lambda . \tag{1}$$

The $\eta$ is then related to the number density $\rho$, mass of a fluid particle $m$, mean speed $\hat{v} = \sqrt{8 k_B T / m \pi}$ of a Boltzmann distribution and mean free path $l_\lambda = 1/(\rho \sigma)$ where $\sigma$ is a cross section. For hard sphere scattering $\sigma = \pi D^2$ for particles with diameter $D$. For this simplified description, the viscosity $\eta$ no longer depends on the number density $\rho$ and is

$$\eta = \frac{1}{3}m\widehat{v}\frac{1}{\pi D^2} = \frac{1}{3}\sqrt{\frac{8}{\pi}}\frac{\sqrt{mk_BT}}{\pi D^2} \quad . \tag{2}$$

This simple result for $\eta$ will be considered as a baseline for comparison. The hard sphere result is modified in several ways:
1. By attractive forces from a more realistic interaction potential.
2. By quantum mechanics especially the unitary limit of quantum gases
3. By the uncertainty principle- $\Delta E \cdot \Delta t \geq \hbar$.

Indications that an energy times time might arise can be seen as follows: $\eta \sim \rho m \widehat{v} l_\lambda$ and writing $l_\lambda \sim \widehat{v} \tau_{coll}$ with $\tau_{coll}$ the time between collisions; then $\eta \sim \rho m \widehat{v}^2 \tau_{coll} \sim \rho E \tau_{coll}$. The product $E\tau_{coll}$ suggests the operation of an uncertainty principle. As mentioned above string theory gave $\eta/s \geq (1/4\pi)(\hbar/k_B)$ where the entropy density varies as $s \sim k_B \rho$, neglecting logarithmic terms for a nucleonic gas. The Sackur-Tetrode expression for the entropy density is $S/V = s = \rho k_B \ln(e^{5/2} g_S / \rho \lambda_T^3)$ for a gas of particles with spin degeneracy $g_S$. The leading order non-logarithmic term is $s \sim (5/2)\rho k_B$.

A perfect liquid has the lowest viscosity allowed by the uncertainty principle. A way of comparing different fluids is given in Ref. [41] where an $\alpha$ parameter is to define as $\eta \equiv \alpha \hbar \rho$. Air has $\eta = 1.8 \cdot 10^{-5} kg/m \cdot s$, $\rho \approx 1.2 kg/m^3 /(30m_p)$ giving a large value of $\alpha = 7500$. Water has $\eta = 10^{-3} kg/m \cdot s$, $\rho \approx 10^3 kg/m^3 /(18m_p)$ and a value $\alpha = 300$. Near perfect fluids have an $\alpha \sim 1$.

___________________________________________________________________

## II.B. Chapman-Enskog Theory

The Chapman-Enskog theory relates the viscosity $\eta$ to either terms involving the classical scattering angle $\chi$ or phase shift $\delta_l$ calculated quantum mechanically from a potential. Fig. 1 shows classical trajectories off various potentials.

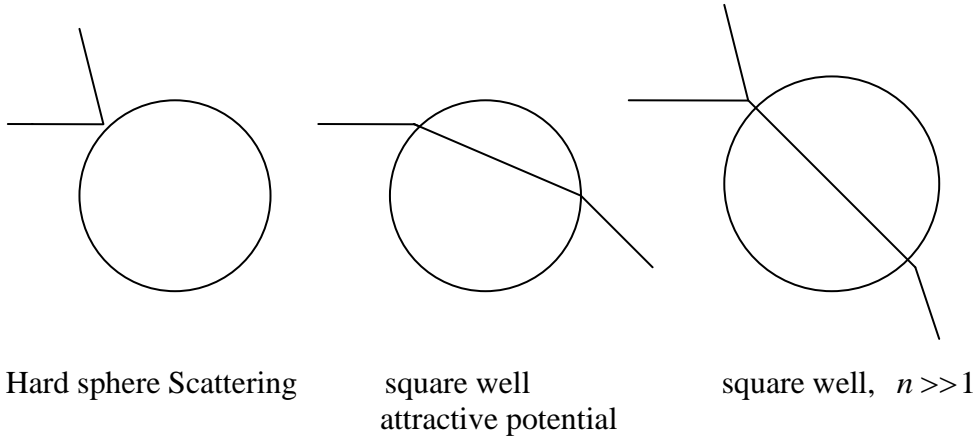

Hard sphere Scattering    square well           square well, $n \gg 1$
                          attractive potential

FIG.1. Classical scattering trajectories off various potentials. Left figure is hard sphere scattering. Middle figure is scattering off an attractive potential. Snell's law of reflection applies to the hard sphere case and gives the scattering angle $\chi$ or angle of deflection with respect to the initial direction as $\chi = \pi - 2\theta_i$. Snell's law of refraction $\sin\theta_i = n\sin\theta_f$ applies to the attractive well and has $\chi = 2(\theta_f - \theta_i)$. Right figure is the scattering off an attractive square well with a very large index of refraction. From this figure it is apparent that a square well with $n \gg 1$ has a viscosity equal to a repulsive hard core potential since the emergent and reflected rays are parallel.

Angular momentum conservation leads to Snell law of refraction which is $n\sin\theta_f = \sin\theta_i$, with $\sin\theta_i = b/R$, and

$$n = n(E, V_0) = \sqrt{(E + |V_0|)/E} = \sqrt{1 + \frac{2\mu|V_0|}{\hbar^2 k^2}} = \sqrt{1 + \frac{2\mu|V_0|}{h^2}\lambda^2} \ . \tag{3}$$

The $b$ is the impact parameter, $E$ is the incident energy and $V_0$ the depth of the potential.

In the Chapman-Enskog approach the viscosity is obtained from

$$\eta = \frac{5}{8}\frac{\sqrt{\pi k_B T m}}{\pi R^2 \omega^{(2,2)}} \tag{4}$$

The $\omega^{(2,2)}$ is evaluated using a Boltzmann weight $e^{-\gamma^2}$ approximation to a Fermi distribution which is a good approximation in dilute systems. The $\omega^{(2,2)}$ is given by

$$\omega^{(2,2)} \equiv \omega = \frac{1}{\pi R^2}\int d\gamma \cdot e^{-\gamma^2}\gamma^7 \phi_{12}^{(2)}(E, V, R) \tag{5}$$

where $\phi_{12}^{(2)}(E,V,R) \equiv \phi$ is evaluated using either a classical scattering angle or using phase shifts from a potential. The $\gamma^2 = E/k_B T$ with $E = \hbar^2 k^2/2\mu$ and with $\mu$ the reduced mass. Table 1 compares the classical theory to the quantum approach. In Table 1 the $\delta_l$ is the $l$'th phase shift of the potential used to describe the scattering. The total cross section is simply $\sigma = (4\pi/k^2)\Sigma_l(2l+1)\sin^2\delta_l$ and this expression has some features that parallel the expression for $\phi$. Difference arise from the $l$ factors and in the argument of the sine functions. For pure $S$-wave scattering $\sigma = \sigma_0 = 4\pi\sin^2\delta_0/k^2$ while $\phi = \phi_0 = (2/3)\sigma_0$.

Table 1 Classical and quantum evaluations of the viscosity functions

| Classical Theory | Quantum Theory |
|---|---|
| $\phi = 2\pi \int_0^\infty (\sin^2 \chi) b \cdot db$ | $\phi = \frac{4\pi}{k^2} \sum_l \frac{(l+1)(l+2)}{2l+3} \sin^2(\delta_{l+2} - \delta_l)$ |
| b=impact parameter | |
| $\chi$ =scattering angle | $\delta_l$ = phase shift for a given potential |
| Snell's laws: | hard sphere potential phase shifts: |
| Hard sphere $\theta_i = \theta_f$ | $\tan \delta_l = j_l(kR)/\eta_l(kR)$ |
| $\chi = \pi - 2\theta_i$ | |
| $\omega^{(2,2)} = 2$ | |
| Attractive potential | $S$ – wave scattering from a square well |
| $\chi = 2(\theta_f - \theta_i)$ | $\delta_0 = \arctan((k/\alpha)\tan \alpha R - kR$, $\alpha^2 = k^2 + \alpha_0^2$ |
| $\sin \theta_i = n \sin \theta_r$ | effective range theory |
| $n = n(E, V_0) = \sqrt{(E + \|V_0\|)/E} =$ | $k \cot \delta_0 = -1/a_{sl} + r_0 k^2 / 2$ |
| $\sqrt{1 + 2\mu V_0 \lambda^2 / h^2}$ | $a_{sl} = R(1 - \tan \alpha_0 R / \alpha_0 R)$, |
| | $r_0 = R - 1/\alpha_0^2 a_{sl} - R^3 / 3a_{sl}^2$, $\alpha_0 = \sqrt{2\mu V_0 / \hbar^2}$ |

The role of spin and statistics is as follows. For identical particles, the $4\pi$ is changed to $8\pi$ in $\phi$ and the sum is over even or odd $l$ states. For particles with spin, spin factors appear. Identical spin $1/2$ fermions interacting through a $S$ – wave, $l = 0$ state are coupled to a total spin zero singlet state. This introduces an additional factor of ¼ in $\phi$. The net effect of both statistics and spin is to reduce $\phi$ by ½, thereby increasing $\eta$ for fermions to twice the value without spin and statistics. For spin 0 bosons, interacting in an $l = 0, S$ – wave, the $4\pi$ is changed to $8\pi$ in $\phi$ and $\eta$ is reduced by ½.

**II. B.1 Classical calculation of viscosity for a hard sphere potential.**

For a hard sphere of radius $R_C$ scattering, the impact parameter $b = R_C \cos \chi / 2$ and $bdb = -(1/4)R_C^2 (\sin \chi) d\chi$. Thus $\phi = 2\pi R_C^2 / 3$ and therefore $\omega^{(2,2)} = 2$. The viscosity is

$$\eta = \frac{5}{16} \frac{\sqrt{\pi k_B T m}}{\pi R_C^2}. \qquad (6)$$

This expression for $\eta$ can be compared to Eq. (1) using $\sigma = \pi R_C^2$. The differential cross section is $\sigma(\chi) = -(b/\sin\chi) \cdot db/d\chi = R_C^2/4$ and total $\sigma = \pi R_C^2$ which is the geometric cross section since anything hitting the sphere is scattered. The ratio of these two expressions for $\eta$ is $(5/16)/(\sqrt{8}/3\pi) = 1.04$, a difference of only 4%.

For hard spheres, a packing fraction defined as $\xi_{pf} = (\pi/6)D^3\rho$ is used in discussions of thermal and transport properties. Specifically, Ref. [42] discuss corrections to $\eta$ obtained from fits of molecular dynamics simulations, which will be used later but noted now, that read

$$\eta = \frac{5}{16} \frac{\sqrt{\pi k_B T m}}{\pi R_C^2}(1.016 + .66000\xi_{pf} + 14.1570\xi_{pf}^2 + 30.82050\xi_{pf}^3 + O(\xi_{pf}^4)) \qquad (7)$$

### II.B.2 Classical calculation of $\eta$ for a square well potential, depth $V_0$, radius $R$

For a square well the refracted angle $\theta_f = \theta_i + \chi/2$, with $\sin\phi_f = \sin\theta_i/n$ and thus

$$\sin\chi = 2\frac{b}{nR}\sqrt{1-\frac{b^2}{n^2R^2}}(1-2\frac{b^2}{R^2}) - 2\frac{b}{R}\sqrt{1-\frac{b^2}{R^2}}(1-2\frac{b^2}{n^2R^2}). \qquad (8)$$

When the index of refraction $n \to 1$, $\chi \to 0$ and when $n \to \infty$, $\theta_f \to 0$, and $\chi \to -2\theta_i$. The differential scattering cross section can be obtained from $\sigma(\chi) = b/\sin\chi |db/d\chi|$. Letting $z = \cos(\chi/2)$, $\sigma(\chi) = n^2R^2(nz-1)(n-z)/(4z(1+n^2-2nz)^2)$. The $\sigma(\chi)$ is constrained by $nz - 1 \geq 0$. The total cross section is $\sigma = \pi R^2$, which is the same $\sigma$ as that of a hard sphere. The viscosity based on Eq. (1) would then have $l_\lambda = 1/\rho\pi R^2$. However, the Chapman-Enskog approach requires an evaluation of $\phi$ and $\omega$ to obtain $\eta$. The $\phi$ is

$$\phi/2\pi R^2 = (16 - 30n - 40n^2 + 20n^3 + 40n^4 + 4n^5 + 20n^7 - 30n^9)/120n^4 +$$

$$(15(n^2-1)(n^8-1)/120n^4)\log[(n+1)/(n-1)]. \qquad (9)$$

The $\phi, \omega \to 0$ if $n \to 1$ and $\eta \to \infty$. However in this limit of infinite viscosity, the concept of momentum transport from collisions between layers of fluid fails since the particles move back and forth between the endpoints defined by the moving walls of the container. For large $n$, the $\phi$ is approximated as $\phi = 2\pi R^2(1/3 - 2/(35n) - 1/(3n^2)...)$. The 1/3 term in the parenthesis gives the hard sphere result. The expression for $\phi$ can be substituted into the integral for $\omega$, using $n = \sqrt{1 + h_T/\gamma^2}$ with $h_T \equiv V_0/k_B T$. This in turn determines $\eta$. The general behavior of $\omega$ with $h_T = V_0/k_B T$ is shown in the right part of Fig. 2.

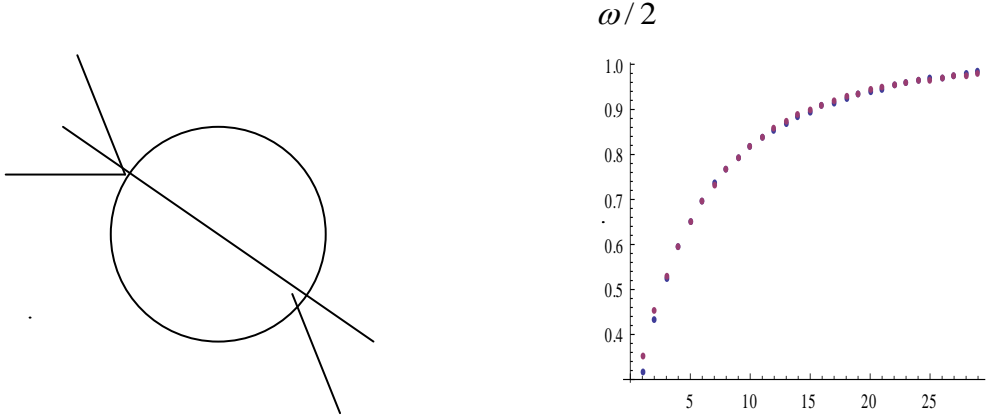

$n \to \infty$ limit of Snell's law: $\theta_r \to 0$
trajectory goes through center & $\omega/2 \to 1$

$h_T = V_0/k_B T$
$\omega/2 \approx (1 - \exp(-0.25(V_0/k_B T)^{0.8}))$

FIG. 2: (Color online) The classical behavior of $\omega/2$ for an attractive potential. Left figure. The left figure is the trajectory which corresponds to $n \to \infty$ and has $\omega/2 \to 1$ which is the hard sphere limit. Right figure. The rise of $\omega/2$ to the value 1 is approximately given by $\omega/2 \approx (1 - \exp(-0.25(h_T)^{0.8}))$. The exponential representation has a slightly higher value at low $h_T$.

Using the results noted in Fig. 2 the viscosity can be approximated as

$$\eta = \frac{5}{16} \frac{\sqrt{\pi k_B T m}}{\pi R^2 (1 - \exp(-0.25(V_0/k_B T)^{0.8}))} \quad . \tag{10}$$

In a classical evaluation of the viscosity, the smallest value of $\eta$ for an attractive interaction is at the largest value of $\omega$ which is the hard sphere result.

The classical approach used Snell's law to develop results involving the viscosity. Snell's law can also be explained by Huygen's wavelets which brings up the next approach based on wave mechanics and the quantum features associated with viscosity.

### II.B.3. Quantum features of viscosity for a hard sphere potential and the semi-classical limit $\hbar \to 0$.

First, the collision between nucleons will be treated as hard sphere scattering off a potential of radius $R_C$. The phase shifts for a hard sphere are given in Table 1 with $j_l$ and

$\eta_l$ Bessel functions. The $\omega$, evaluated in the Boltzmann limit, can be rewritten as

$$\omega = 4\xi^4 \int_0^\infty dx \cdot e^{-\xi x^2} x^7 \left( \frac{1}{x^2} \sum_{l=0,1,2,\ldots} \frac{(l+1)(l+2)}{2l+3} \sin^2(\delta_{l+2}(x) - \delta_l(x)) \right) \quad (11)$$

The $x = kR_C$, $\xi = (\lambda_T / R_C \sqrt{2\pi})^2$ and $k$ is the wave number. The quantum wavelength $\lambda_T = h / \sqrt{2\pi k_B T m}$. The $\phi$ sum has the following scaling property when $x \to \infty$ which is $\phi \to 2\pi R_C^2 / 3$ and is the classical hard sphere result mentioned above. The scaling behavior of $\phi(x)$ is shown in left side of Fig. 3. This scaling result parallels a similar result for the cross section which, in the high energy limit, is $\sigma \to 2\pi R_C^2$. The factor of 2 increase over the geometrical area $\pi R_C^2$ arises from Fraunhofer diffraction [43].

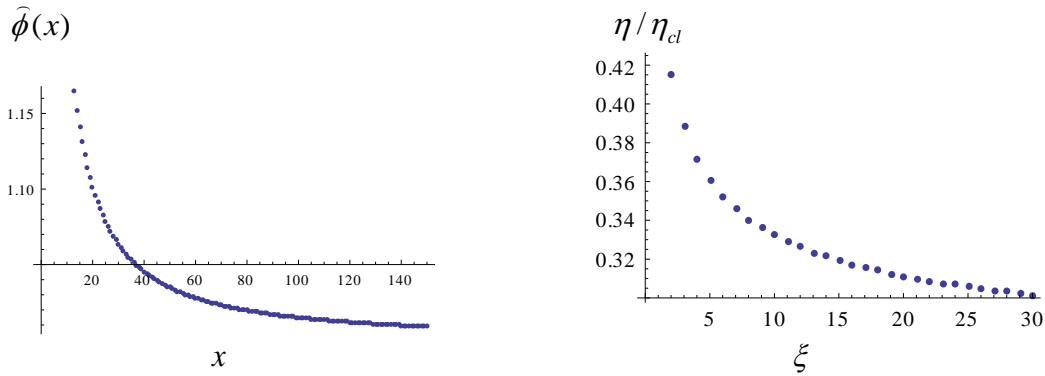

FIG. 3. Left Figure. Scaling property of $\hat{\phi}(x)$ with $x$. The quantity plotted is $\hat{\phi} \equiv (6/x^2)\Sigma_l[(l+1)(l+2)/(2l+3)]\cdot\sin^2(\delta_{l+2}(x) - \delta_l(x))$ versus $x$. For this rescaled quantity the limiting value is unity. The $\phi = 2\pi R_C^2 \hat{\phi}/3$. Right Figure. The hard sphere quantum viscosity as a function of $\xi$. The vertical axis is $\eta/\eta_{cl}$, the ratio of the quantum result for the viscosity divided by the classical hard sphere result. The latter is given by Eq. (6) with $\omega = 2$. At low values of $\xi$, the quantum calculation is the same as the classical result and $\eta/\eta_{CL} \to 1$. At large $\xi$ the ratio $\eta/\eta_{CL} = \eta_{S-wave}/\eta \to 1/4$.

At very low energies only an $S$−wave hard sphere phase shift is important in the sum in Eq. (11). The cross section goes to $4\pi R_C^2$, or four times the geometrical result. Also, $\phi \to 8\pi R_C^2 / 3$ for $x << 1$. The $S$−wave phase shift is $\delta_0 = -kR_C = -x$. Using $\sin^2 x \approx x^2$, the resulting $\omega = 8$, which is 4 times the classical value $\omega = 2$. The range of $\eta$ is then

$$\frac{5}{16}\frac{\sqrt{\pi k_B T m}}{\pi R_C^2} \geq \eta \geq \frac{5}{64}\frac{\sqrt{\pi k_B T m}}{\pi R_C^2} \ . \tag{12}$$

The left hand side of Eq. (12) is the quantum scaling limit result and is the same as the classical hard sphere result for $\eta$. The right hand side of Eq. (12) is the pure $S-$wave hard sphere scattering result. The hard sphere quantum result between these two limits for $\eta$ as a function of $\xi$ is shown in the right part of Fig 3. The behavior of $\omega$ is determined by $\xi = (\lambda_T / R_C \sqrt{2\pi})^2$ which in turn depends on the temperature $T$ through $\lambda_T$. The semi-classical limit has $\hbar \to 0$ and $\xi \to 0$. In this limit, the scaling behavior shown in the right side of Fig. 3 arises and $\eta$ reaches the classical value. To see how $\eta$ evolves from the pure $S-$wave limit to the classical value, small $x$ expansions are made for the $P, D, F$ phase shifts: $\delta_1 = -x^3/3 + x^5/5 - x^7/7$, $\delta_2 = -x^5/45 + x^7/189$, $\delta_3 = -x^7/1575$, to order $x^7$. The resulting $\omega$, further expanded in even $l$ and odd $l$ components $\omega_E$ and $\omega_O$, is

$$\omega = \left(8|_S - \frac{32}{3\xi}|_S + \frac{1}{\xi^3}(\frac{360}{7}|_{S,D} + \frac{256}{35}|_D)..\right)_E + \left(\frac{48}{\xi^2}|_P - \frac{1152}{5\xi^3}|_P + ..\right)_O \equiv \omega_E + \omega_O. \tag{13}$$

The factor $8$ is the pure $S-$wave result. The $1/\xi^2$ term arises solely from a $P-$wave. The contribution of each partial wave is also given. It should be noted that the result of Eq. (13) gives an expansion for $\omega$ in inverse powers of $\hbar^2$ since $\xi = (\lambda_T / R_C \sqrt{2\pi})^2 \sim \hbar^2$. The viscosity is connected to this series expansion around the $S-$wave scattering limit using Eq. (4). The hard sphere quantum result for $\eta$ is shown in Fig. 3 as a function of $\xi$.

**II.B.4 Quantum aspects of viscosity for a square well potential and the unitary limit**

For a square well potential $\tan \delta_l = \{kj_l'(x) - \gamma_l j_l(x)\} / \{kn_l'(x) - \gamma_l n_l(x)\}$ with $\gamma_l = \alpha j_l'(y) / j_l(y)$ [43]. The $\alpha = \sqrt{k^2 + |V_0| 2\mu/\hbar^2}$, $y = \alpha R$, $x = kR$. The prime superscripts represent derivatives with respect to $x$ or $y$. The $y = n(E, V_0) x$, with $n(E, V_0) = \sqrt{1 + |V_0|/E}$ the index of refraction of the classical description. The low energy behavior will be considered as a baseline for comparison. In this limit the $S-D$ wave phase shift $\delta_2 - \delta_0 \approx -\delta_0$ with $\delta_0 = \arctan[(kR/\alpha R) \tan \alpha R - kR$. An effective range approximation for $\delta_0$ reads $k \cot \delta_0 = -1/a_{sl} + r_0 k^2 / 2$. The scattering length $a_{sl} = R(1 - \tan \alpha_0 R / \alpha_0 R)$ and the effective range is $r_0 = R - 1/\alpha_0^2 a_{sl} - R^3 / 3 a_{sl}^2$. The $\alpha_0 = \sqrt{2\mu V_0 / \hbar^2}$. For large $a_{sl}$ the $r_0 \approx R$. A zero energy bound state appears when $\alpha_0 R = \pi/2$. Then $a_{sl} \to \infty$. Similarly, for a zero energy resonant like state the $a_{sl} \to -\infty$. The $\omega$ is

$$\omega = 4\xi^4 \int_0^\infty dx \cdot e^{-\xi x^2} x^7 \frac{1}{x^2} \left( \frac{(ka_{sl})^2}{1 + a_{sl}(a_{sl} - r_0)k^2 + (a_{sl}(a_{sl} - r_0)k^2)^2 \frac{r_0^2}{4(a_{sl} - r_0)^2}} \right) \quad (14)$$

in an effective range approximation and in a Boltzmann limit. When $|a_{sl}| \gg r_0$, then

$$\omega = 4\frac{2}{3}\xi \left( \frac{2 + \zeta(\zeta - 1) - e^\zeta \zeta^3 \Gamma(0, \zeta)}{2} \right) \quad (15)$$

The $\Gamma(0, g) = E_1(g) = \int_g^\infty e^{-t} t^{-1} dt$ and $\zeta = \xi R^2 / a_{sl}^2 = (\lambda_T / a_{sl} \sqrt{2\pi})^2$. The limit $|a_{sl}| \to \infty$ is the unitary limit. In the unitary limit $\omega \to 8\xi/3 = 8(\lambda_T / R\sqrt{2\pi})^2 / 3$. Thus $\omega$ introduces quantum effects via the factor $\lambda_T$. The $S$-wave unitary or universal thermodynamic limit for $\eta$ is determined by the quantum wavelength $\lambda_T$ which reads

$$\eta \to \frac{15}{32} \frac{\sqrt{\pi k_B T m}}{\lambda_T^2} = \frac{15}{16} \frac{(\pi k_B T m)^{3/2}}{h^2} = \frac{15}{32} \sqrt{2\pi} \frac{\hbar}{\lambda_T^3}. \quad (16)$$

The last equality in Eq. (16) shows that the viscosity is proportional to Planck's constant divided by the quantum volume $\lambda_T^3$. By contrast the $S$-wave hard sphere limit can be written as $\eta = (5\sqrt{2\pi}/16) \cdot \hbar/(\lambda_T \pi R_C^2)$ but is $\hbar$ independent.

As noted above, for identical particles, the $4\pi$ is changed to $8\pi$ in $\phi$ and the sum is over even or odd $l$ states. Identical spin $1/2$ fermions interacting through an $S$-wave, $l = 0$ state are coupled to a total spin zero singlet state so that the total wavefunction is anti-symmetric. This introduces an additional factor of ¼ in $\phi$. The net effect is to reduce $\phi$ by ½, thereby increasing $\eta$ for fermions to

$$\eta = \frac{15}{16} \sqrt{2\pi} \frac{\hbar}{\lambda_T^3}. \quad (17)$$

The unitary limit for $\eta$ is independent of the potential used since it is based on an effective range result and with $a_{sl} \to \infty$. A calculation of $\eta$ with a delta shell potential [30] gave the same result and also the same result can be found in Ref. [26,27]. The result of Eq.(17) is also noted in ref. [36]. In the next subsection, a very accurate approximation to the viscosity developed from Eq. (15) will be given which contains corrections away from the unitary limit and also corrections from the effective range.

## II.B.6 Low energy behavior of the viscosity of a dilute neutron gas

The viscosity of a dilute gas of neutrons will now be considered in more detail. Previous studies [22-23] of the second virial coefficient showed that the $S-$ wave approximation accurately described the scattering up to temperatures of ~15 $MeV$ before $P-$ wave and $D-$ wave contributions start to become significant. In a space symmetric $l=0$ $S-$ wave state, the neutrons are coupled to a total spin $\vec{S}=0$ antisymmetric state. In this channel the observed $S-$ wave scattering length is $a_{sl}=-17.4\,fm$ and the effective range is $r_0=2.4\,fm$. The $\omega$ integral of Eq. (15), when corrected for an effective range contribution, leads to a viscosity

$$\eta = \frac{(a_{sl}(a_{sl}-r_0))}{a_{sl}^2}\left[\frac{2}{2+\zeta(\zeta-1)-\zeta^3 e^\zeta \Gamma(0,\zeta)}\right]\left(\frac{15}{16}\sqrt{\pi}\,\frac{\hbar}{\lambda_T^3}\right) \qquad (18)$$

with $\zeta = \lambda_T^2/(2\pi a_{sl}(a_{sl}-r_0)) = (\hbar c)^2/(mc(a_{sl}(a_{sl}-r_0)k_B T) = 0.12/(k_B T/1MeV)$. The factor $(a_{sl}-r_0)/a_{sl} = -19.8/(-17.4) = 1.138$. The last curved bracket term in Eq. (18) is the unitary limit. In the above equation the factor in square bracket is reasonably approximated by $1+\zeta/3$ for the entire range of $\zeta$. The $\zeta/3$ is obtained from the asymptotic value of $2+\zeta(\zeta-1)-\zeta^3 e^\zeta \Gamma(0,\zeta)$ for large $\zeta$. The value of 1 in $1+\zeta/3$ is the unitary limit. Thus the viscosity is can be approximated by

$$\eta = \left(\frac{a_{sl}(a_{sl}-r_0)}{a_{sl}^2} + \frac{\lambda_T^2}{3\cdot 2\pi a_{sl}^2}\right)\left(\frac{15}{16}\sqrt{2\pi}\,\frac{\hbar}{\lambda_T^3}\right). \qquad (19)$$

The first term alone in the first bracket is somewhat larger than the unitary limit since $a_{sl}(a_{sl}-r_0)/a_{sl}^2 = (|a_{sl}|+r_0)/|a_{sl}|>1$ for neutrons. The first term when combined with the factor $\hbar/\lambda_T^3$ contains Planck's constant. The second term, which involves $\lambda_T^2/a_{sl}^2$, when combined with the factor $\hbar/\lambda_T^3$, is independent of Planck's constant. Thus the expression of Eq. (19) contains both $\hbar$ dependent quantum aspects and $\hbar$ independent semiclassical aspects.

The ratio of viscosity to entropy density $\eta/s \sim (\hbar/k_B)(1/(\lambda_T^3\rho)$ brings in the factor $1/(\rho\lambda_T^3)$, which is related to the fugacity. Low values of $\eta/s \sim (\hbar/k_B)(1/(\lambda_T^3\rho)$ occur at high $(\rho\lambda_T^3)$. At high $(\rho\lambda_T^3) \sim 1$, higher order terms are important in both the viscosity and entropy density. In the unitary limit

$$\eta = \left(\frac{a_{sl}(a_{sl}-r_0)}{a_{sl}^2}\right)\left(\frac{15}{16}\sqrt{2\pi}\right)\left(\frac{1}{\rho\lambda_T^3}\right)\hbar\rho \qquad (20)$$

which is of the form $\eta \equiv \alpha\hbar\rho$ suggested in ref. [41] but with $\alpha$ dependent on the factor $(\rho\lambda_T^3)$. For a rough estimate a value $(\rho\lambda_T^3)=1/2$ is used. Then $\alpha\sqrt{2\pi}\,2\sim 9$, a value suggesting a system which may not be too far from being a nearly fluid, but a better development of higher order terms is necessary. The importance of higher order corrections is partly discussed in the next section for a hard sphere gas.

It should be noted that if a system had a bound state then $a_{sl} > 0$ and $a_{sl}(a_{sl} - r_0)/a_{sl}^2 < 1$. In this case the viscosity is lower than a quasi resonant state.

### II.B.7. Viscosity to entropy density ratio from molecular dynamic studies.

The results of Ref. [42] can be used to make a qualitative estimate of the viscosity to entropy density ratio based on molecular dynamic studies. The viscosity expression was already given in Eq. (7). Ref. [42] also contains an equation of state which reads

$$P = \rho k_B T \frac{1 + \xi_{pf} + \xi_{pf}^2 - \xi_{pf}^3}{(1 - \xi_{pf})^3} \quad (21)$$

The associated interaction entropy can be found from the Maxwell relation $(\partial P / \partial T)_V = (\partial S / \partial V)_T$. Including the Sackur-Tetrode entropy leads to

$$s = \frac{S}{V} = \rho k_B \left( Ln[e^{5/2} \frac{1}{\rho \lambda_T^3}] - 4 \frac{\xi_{pf}(1 - (3/4)\xi_{pf})}{(1 - \xi_{pf})^2} \right) \quad (22)$$

Minimizing $\eta / s$ with respect to $T$ gives a minimium $T$ determined by $z_m \equiv \rho \lambda_{T_m}^3 = 1/\sqrt{e}$ where $\lambda_{T_m} = \hbar \sqrt{2\pi / m k_B T_m}$. The resulting $\eta / s$ is then

$$\frac{\eta}{s} = \frac{5\sqrt{2} e^{1/6}}{48} \frac{1 + .66 \xi_{pf} + 14.157 \xi_{pf}^2 + 30.820 \xi_{pf}^3}{(\frac{6}{\pi})^{2/3} \xi_{pf}^{2/3} (1 - \frac{4}{3} \xi_{pf} \frac{1 - (3/4)\xi_{pf}}{(1 - \xi_{pf})^2})} \frac{\hbar}{k_B} \quad (23)$$

The minimum in $\eta / s$ with packing fraction $\xi_{pf}$ occurs at $\xi_{pf} = 0.11$. At $\xi_{pf} = 0.11$ then $\eta / s = 0.8 \hbar / k_B$ which is about 10 times the minimum $\eta / s = \hbar /(4\pi k_B)$. Note that the viscosity is purely classical but the quantum aspects in the $\eta / s$ ratio arise from the quantum aspects of the entropy from the Sackur-Tetrode law.

## III. Conclusions and summary

Viscosity is important in many areas of physics. For example, in low energy to relativistic energy heavy ion collisions viscosity affects the collective flow of the fluid matter. As noted current interest arose from string theory and questions related to perfect fluid behavior. String theory gave a minimum value for the viscosity to entropy density ratio which is connected to an uncertainty principle. One can ask the following questions. 1. What is the viscosity of a nucleonic fluid? 2. How perfect is it? In an attempt to answer question 1, the viscosity was studied in both a classical and a quantum approach for several types of potentials. The potentials included treating the collisions between nucleons as: A) billiard ball hard spheres scattering, B) interactions represented by an attractive square well. Both A) and B) are treated classically and quantum mechanically. The billard ball classical model is used as a baseline for comparison with the other cases considered. One comparison showed that the smallest classical value of the viscosity for an attractive potential is just the hard sphere limit. This result is also easily seen (see Fig. 1&2) when the attractive potential result was cast into Snell's law of reflection and refraction with an energy dependent and potential dependent index of refraction. The hard sphere quantum result for the viscosity was shown to reduce to the hard sphere classical result in a scaling limit of the quantum approach (see Fig. 3). Specifically, the short wavelength limit of the quantum result have scaling laws associated with it which are similar to those associated with the Fraunhoher diffraction increase for the hard sphere geometric cross section. The hard sphere quantum result for the viscosity was shown to vary between the hard sphere classical limit and ¼ this value, with the latter arising from pure $S-$wave scattering in the low energy limit of the theory. The ¼ factor parallels the $S-$wave total cross section which is 4 times the geometric hard sphere cross section. The viscosity in a quantum approach for an attractive square well was also developed in the unitary limit using an effective range theory. The unitary limit arises when the scattering length is infinite. In the unitary limit, Planck's constant explicitly appears in the viscosity. An expression was also given (Eq. (19) (a more exact expression is given in Eq. (18)) which interpolates between the unitary limit and regions away from the unitary limit in which the viscosity is independent of Planck's constant, but dependent on the scattering length. Effective range corrections to the theory were also developed. In the case of a system of pure neutrons the unitary limit can be realized in a certain $\rho$ and $T$ range. For neutrons the $S-$wave scattering length is $a_{sl} = -17.4\,fm$ and the effective range is $r_0 = 2.4\,fm.$ Question 2 presented problems because the minimum value of $\eta/s$ occurs in regions were higher order interaction corrections are important. Some observations were presented regarding the near perfect fluid behavior of a system of nucleons. These observations came from molecular dynamics studies of hard sphere gases which included packing fraction corrections. In particular $\eta/s \sim \hbar/k_B$ was shown to arise from a classical description of the viscosity but quantum description of the entropy density. A numerical study gave a result about 10 times the minimum $\eta/s \geq \hbar/(4\pi k_B)$, suggesting a system somewhat close to being a nearly perfect fluid.

This work is supported by Department of Energy under Grant DE-FG02ER-409DOE
* Work done in part at the Department of Physics, California Institute of Technology, Pasadena, Ca, 91125

1515